\documentclass[nofootinbib,pra,superscriptaddress,a4paper,twocolumn]{revtex4-1}
\usepackage{geometry}
\usepackage[english]{babel}
\usepackage[latin1]{inputenc}
\usepackage{hyperref}
\usepackage{amsfonts}
\usepackage{amsmath, amssymb, amsfonts, mathrsfs}
\usepackage{graphicx}
\usepackage{xspace}
\usepackage{tcolorbox}
\usepackage{natbib}
\usepackage{multirow}
\usepackage[bottom]{footmisc}
\usepackage{array}
\usepackage{tabulary}

\newcommand{\ket}[1]{|#1\rangle}
\newcommand{\bra}[1]{ \langle #1 \,  |}

\newcommand*{\melvin}{{\small M}{\scriptsize ELVIN}\xspace}
\newcolumntype{P}[1]{>{\centering\arraybackslash}p{#1}}

\begin{document}

\title{Quantum Experiments and Graphs III:\\ High-Dimensional and Multi-Particle Entanglement}

\author{Xuemei Gu}
\email{xmgu@smail.nju.edu.cn}
\affiliation{State Key Laboratory for Novel Software Technology, Nanjing University, 163 Xianlin Avenue, Qixia District, 210023, Nanjing City, China.}
\affiliation{Institute for Quantum Optics and Quantum Information (IQOQI), Austrian Academy of Sciences, Boltzmanngasse 3, 1090 Vienna, Austria.}

\author{Lijun Chen}
\email{chenlj@nju.edu.cn}
\affiliation{State Key Laboratory for Novel Software Technology, Nanjing University, 163 Xianlin Avenue, Qixia District, 210023, Nanjing City, China.}

\author{Anton Zeilinger}
\email{anton.zeilinger@univie.ac.at}
\affiliation{Institute for Quantum Optics and Quantum Information (IQOQI), Austrian Academy of Sciences, Boltzmanngasse 3, 1090 Vienna, Austria.}
\affiliation{Vienna Center for Quantum Science \& Technology (VCQ), Faculty of Physics, University of Vienna, Boltzmanngasse 5, 1090 Vienna, Austria.}

\author{Mario Krenn}
\email{mario.krenn@univie.ac.at; present address: Department of Chemistry, University of Toronto, Toronto, Ontario M5S 3H6, Canada.}
\affiliation{Institute for Quantum Optics and Quantum Information (IQOQI), Austrian Academy of Sciences, Boltzmanngasse 3, 1090 Vienna, Austria.}
\affiliation{Vienna Center for Quantum Science \& Technology (VCQ), Faculty of Physics, University of Vienna, Boltzmanngasse 5, 1090 Vienna, Austria.}

\begin{abstract}
Quantum entanglement plays an important role in quantum information processes, such as quantum computation and quantum communication. Experiments in laboratories are unquestionably crucial to increase our understanding of quantum systems and inspire new insights into future applications.  However, there are no general recipes for the creation of arbitrary quantum states with many particles entangled in high dimensions. Here, we exploit a recent connection between quantum experiments and graph theory and answer this question for a plethora of classes of entangled states. We find experimental setups for Greenberger-Horne-Zeilinger states, W states, general Dicke states, and asymmetrically high-dimensional multipartite entangled states. This result sheds light on the producibility of arbitrary quantum states using photonic technology with probabilistic pair sources and allows us to understand the underlying technological and fundamental properties of entanglement.
\end{abstract}
\date{\today}
\maketitle
Entanglement, which exhibits correlations without a classically analog \cite{einstein1935can, bell1964einstein}, is a very peculiar property of quantum states. It is of particular importance in understanding the foundations of quantum mechanics, especially for local realism. Nowadays it has been viewed as a prominently useful resource for quantum information applications, such as quantum computation and quantum communication.

The smallest entangled system consists of two particles, which share one bit of information (such as the polarization state of a photon) in a non-local-realistic way. Such a system is a cornerstone of research in quantum entanglement theory.

More particles or high-dimensional degrees of freedom can lead to more complex types of entanglement. A prominent example of multipartite entanglement is the so-called Greenberger-Horne-Zeilinger (GHZ) state \cite{greenberger1989going, greenberger1990bell}, which offers a new understanding in the study of our local and realistic worldview. Another famous class of entangled states is the Dicke state \cite{dicke1954coherence}, with an important special case -- the W state.

Increasing the number of involved degrees of freedom in the entanglement significantly increases the number of different possible states and the complexity of studying them. For example, the question about \textit{all-versus-nothing} violations of high-dimensional GHZ states has only been understood in 2014 \cite{ryu2014multisetting, lawrence2014rotational}, and these states have only been experimentally implemented in the very recent past \cite{erhard2018experimental}.  High-dimensional and multipartite entanglement can lead to new, asymmetric types of quantum correlations which are not seen in any qubit system \cite{huber2013structure, goyeneche2016multipartite}. Such a type of entanglement was first been investigated in the laboratory in 2016 \cite{malik2016multi} and allows potentially different types of quantum communication scenarios \cite{pivoluska2018layered}.

In the spirit of Richard Feynman, who once famously said \textit{"What I cannot create, I do not understand,"} here we ask, \textit{Which quantum entangled states can be created in the laboratories with current photonic technologies?}

\begin{table}[!t]
  \centering
  \caption{The analogies between graph theory and quantum experiments.}
    \begin{tabular}{p{3.3cm}|p{4cm}}
    \hline
    \textbf{Graph Theory}&\textbf{Quantum Experiments}\\ \hline
    undirected Graph& optical setup with nonlinear crystals\\ \hline
    Vertex&optical output path\\ \hline
    Edge&nonlinear crystal\\ \hline
    colors of the edge &mode numbers \\ \hline
    perfect matching & $n$-fold coincidence \\ \hline
    \#(perfect matchings)&\#(terms in quantum state)\\ \hline
    \end{tabular}
   \label{tab:compare}
\end{table}
Using a recently uncovered bridge between quantum experiments with probabilistic photon pair sources and graph theory \cite{krenn2017quantum}, we answer this question for many classes of entangled states. The correspondence is listed in Table \ref{tab:compare}. Our strategy is to translate the question about the construction of a quantum state into a question about the existence of a graph with certain properties. All of our affirmative answers are constructive, meaning that in these cases we show the graph and its corresponding quantum experimental setup.

In this paper, we briefly summarize the main results from \cite{krenn2017quantum} and explain the connection between quantum experiments and graphs. Then we show graphs and experimental setups for creating 2-dimensional and 3-dimensional GHZ states as well as 4-particle W state. Afterwards, we extend the applications and find a construction for W state with arbitrary particles, and its generalization -- the Dicke states. Furthermore, we present a general solution to producing high-dimensional 3-particle entangled states, which answers a question that has been raised more than 3 years ago.

Our investigation significantly enlarges the understanding of currently existing experimental technology and finds systematic solutions to a question that has previously investigated only with advanced automated search methods \cite{krenn2016automated, melnikov2018active}.

\section*{Generation of Greenberger-Horne-Zeilinger states}
GHZ states form a very important class of entangled states and are denoted as
\begin{equation}
\ket{GHZ_{n}^{d}}=\frac{1}{\sqrt{d}}\sum_{i=0}^{d-1} \ket{i}^{\bigotimes n}
\end{equation}
where $n$ is the number of particles and $d$ is the dimension for every particle.

\begin{figure}[!t]
\centering
\includegraphics[width=0.45\textwidth]{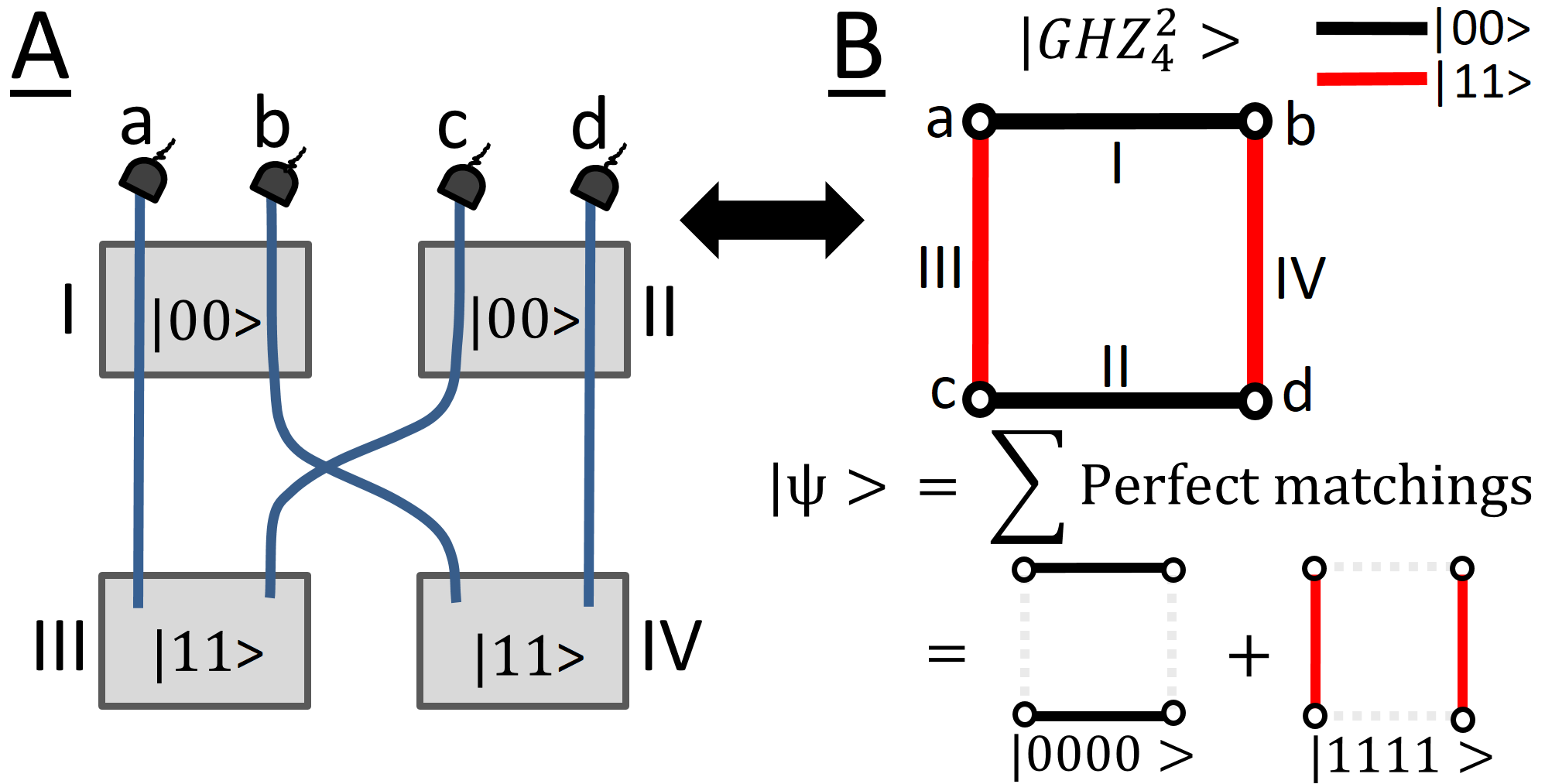}
\caption{Experiment for producing a 2-dimensional 4-particle GHZ state $\ket{GHZ_{4}^{2}}$ based on \textit{Entanglement by Path Identity} \cite{krenn2017entanglement} and corresponding graph \cite{krenn2017quantum}. \textbf{A}: Four nonlinear crystals (gray squares) are pumped coherently and the pump laser can be set such that two photon pairs are created with reasonable probabilities. The final quantum state is created conditionally on simultaneously clicks in all four detectors. \textbf{B}: In the graph, every vertex stands for a photon's path and every edge represents a nonlinear crystal. The color depicts the mode number of a photon. Here black and red [dark gray] edges correspond to state with $\ket{00}$ and $\ket{11}$, respectively. A four-fold coincidence in the experiment can be seen as a subset of edges that contains every vertex only once, which is called as a perfect matching in the graph. Thus, the coherent superposition of two perfect matchings leads to four-fold coincidences, which describes the quantum state $\ket{\psi}_{abcd}=\frac{1}{\sqrt{2}}(\ket{0000}+\ket{1111})$.}
\label{fig:fig1GHZ2d}
\end{figure}

In Fig. \ref{fig:fig1GHZ2d}A, we show an experimental setup to produce a 2-dimensional 4-particle GHZ state $\ket{GHZ_{4}^{2}}$ using \textit{Entanglement by Path Identity} \cite{krenn2017entanglement}. Photon pairs can be created by probabilistic photon pair sources (such as nonlinear crystals, depicted as gray squares) via the spontaneous parametric down-conversion (SPDC) process. The crystals are set up in such a way that crystals I and II produce photons with states $\ket{00}$, while crystals III and IV produce photons with states $\ket{11}$. Here the mode numbers $0$ and $1$ correspond to the polarization of photons\footnote{A photon's mode numbers can be changed by inserting variable mode-shifters in the photon's paths. For convenience, we neglect the mode-shifters and label the mode numbers in the nonlinear crystal.}, the orbital angular momentum (OAM) \cite{allen1992orbital, yao2011orbital, krenn2017orbital} or some other degree-of-freedom such as time-bin \cite{franson1989bell, versteegh2015single} or frequency \cite{olislager2010frequency}.

The four crystals are pumped coherently and the pump power is set in such a way that two photon pairs are produced with reasonable probabilities\footnote{A higher number of photon pairs can be created in the down-conversion process. However, one can adjust the laser power such that these cases have a sufficiently low probability, which can be neglected.}. In the experiment, the final quantum state is obtained by post-selection on \textit{4-fold coincidences, which means that all $4$ detectors click simultaneously}. This happens when two photon pairs origin either from crystals I and II or from crystals III and IV. No other event could contribute to the 4-photon coincidences. For example, if only the photon pairs are produced from crystals II and III, there will be two photons in path $c$ and no photon in path $b$.
\begin{figure*}[!t]
\centering
\includegraphics[width=0.98\textwidth]{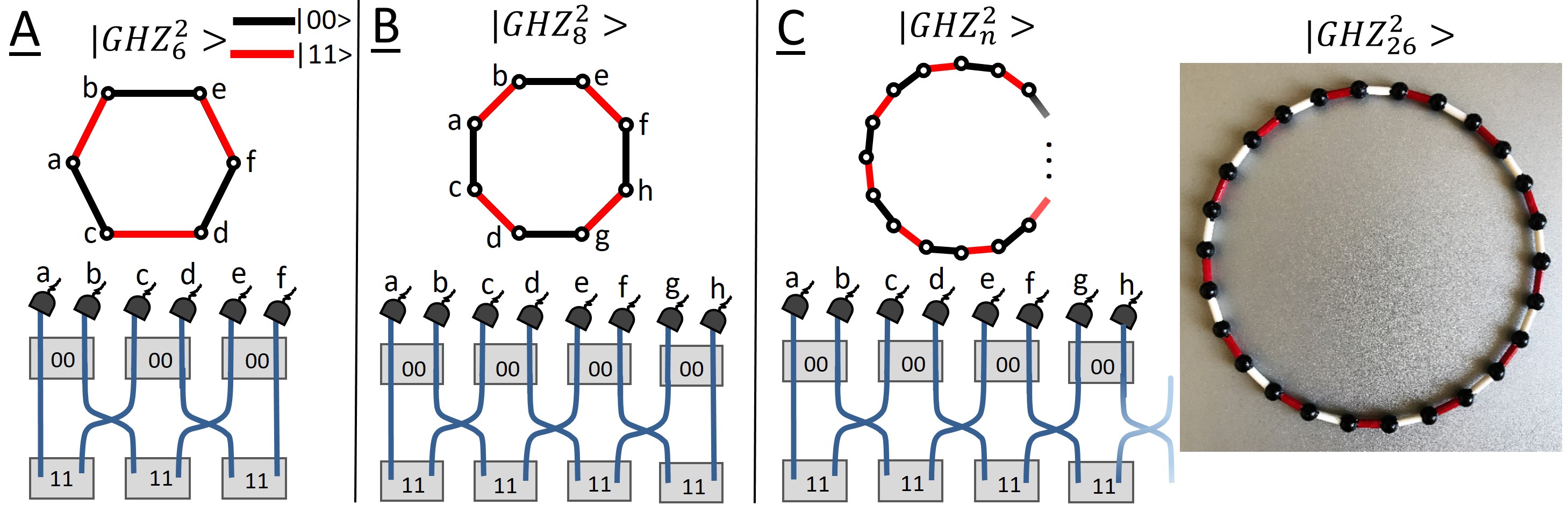}
\caption{General graphs and experimental implementations for creating 2-dimensional n-particle GHZ states $\ket{GHZ_{n}^{2}}$. In Fig. \ref{fig:fig1GHZ2d}B, we have shown a graph for a 2-dimensional 4-particle GHZ state $\ket{GHZ_{4}^{2}}$. One can arbitrarily extend that graph, which means one can arbitrarily increase the number of particles in the quantum state. \textbf{A}, \textbf{B} and \textbf{C} show the general graphs and experiments for producing 2-dimensional 6, 8 and $n$-particle GHZ states. On the right side, we also show a 3D printed graph, which corresponds to a 2-dimensional 26-particle GHZ state $\ket{GHZ_{26}^{2}}$. There the mode numbers $0$ and $1$ are represented with white and red [dark gray] colors, and the vertices are depicted in black.}
\label{fig:fig2GHZ2dn}
\end{figure*}

One can translate such an optical setup into a graph \cite{krenn2017quantum}, which is described in Fig. \ref{fig:fig1GHZ2d}B. There the vertices depict the photon's paths and the edges represent the nonlinear crystals. The graph contains two subsets of edges $(E_{ab}, E_{cd})$ and $(E_{ac}, E_{bd})$. Each subset contains \textit{all four vertices only once, which is called as a perfect matching of the graph}. Therefore, the four-fold coincidences in the experiment are given by the coherent superposition of perfect matchings of the graph. The quantum state after conditioning on four-fold coincidences can be written as
\begin{equation}
\ket{\psi}_{abcd}=\frac{1}{\sqrt{2}}(\ket{0000}+\ket{1111})
\end{equation}
where values $0$ and $1$ stand for photon's mode numbers (such as the OAM modes of the photon), and the subscript $a$, $b$, $c$ and $d$ represent the photon's paths.

Now we generalize this technique to 2-dimensional n-particle GHZ states $\ket{GHZ_{n}^{2}}$. One can arbitrarily increase the number of vertices of the graph in Fig. \ref{fig:fig1GHZ2d}B, which means that the number\footnote{A probabilistic photon pair source (such as a nonlinear crystal) produces photon pairs, thus the number of particles $n$ is an even number. However, some of the photons can be seen as triggers such that the number $n$ can be an odd number.} of particles can be arbitrarily large. We show the general graphs and experiments for creating 2-dimensional $n$-particle GHZ states $\ket{GHZ_{n}^{2}}$ in Fig. \ref{fig:fig2GHZ2dn}. These graphs can describe, for instance, the largest polarization GHZ state consisting of $n=12$ photons \cite{zhong201812}. \footnote{Interestingly, the largest GHZ state ever produced in any platform is an 18-qubit state encoded in three degrees of freedom with six photons \cite{wang201818}. It would be interesting to extend the current graph language to cover such hyper-entangled multiphotonic quantum states.}

\begin{figure}[!t]
\centering
\includegraphics[width=0.47\textwidth]{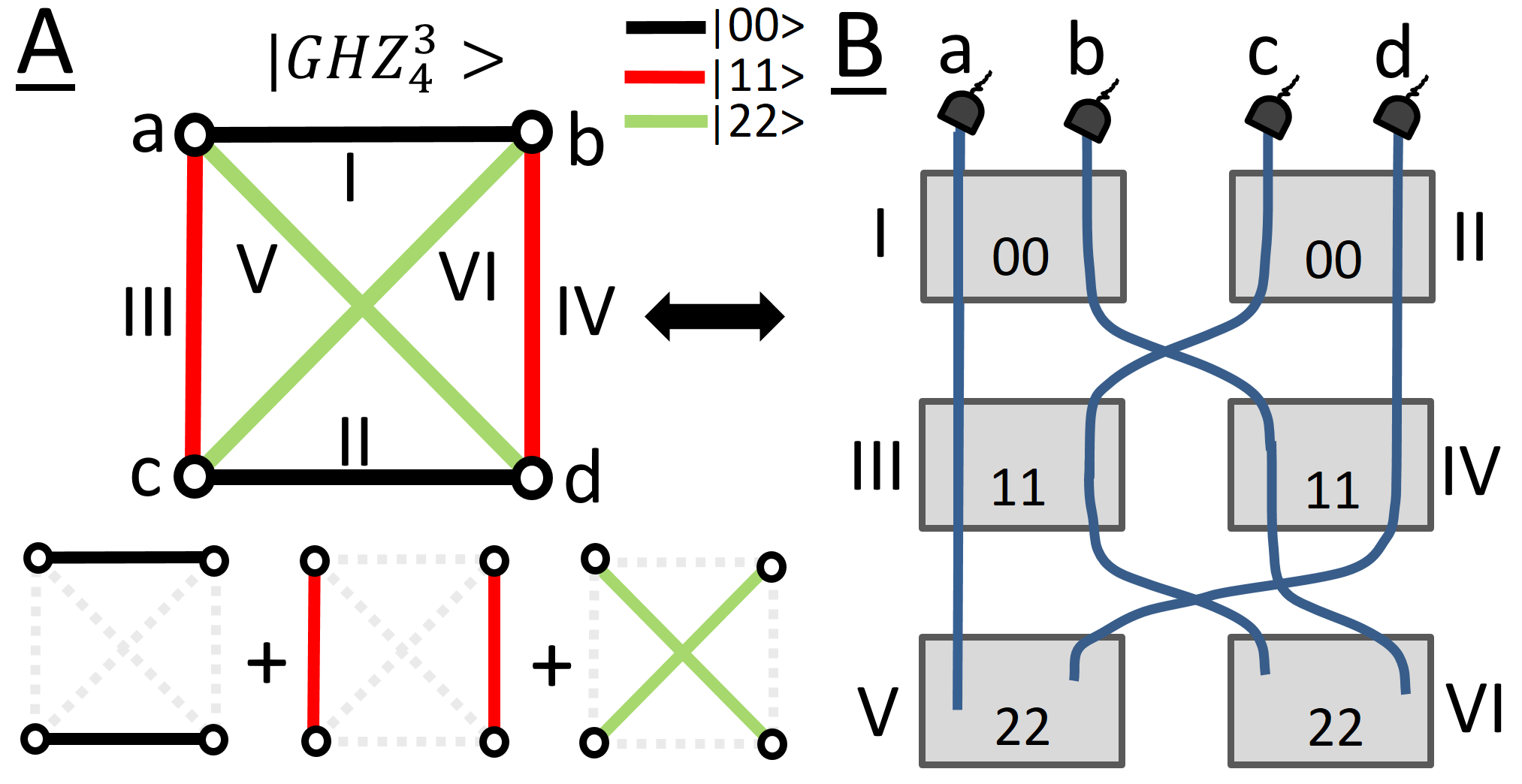}
\caption{Graph and optical setup for producing a 3-dimensional 4-particle GHZ state $\ket{GHZ_{4}^{3}}$. \textbf{A}: We add one perfect matching of the graph in Fig. \ref{fig:fig1GHZ2d}B. The black, red [dark gray] and green [light gray] edges stand for the corresponding crystals producing the photon pairs with states $\ket{00}$, $\ket{11}$ and $\ket{22}$, respectively. \textbf{B}: The corresponding experimental setup of the graph. All crystals are pumped coherently and the laser power can be set such that two photon pairs are produced. The coherent superposition of three perfect matchings leads to the quantum state, which is $\ket{\psi}_{abcd}=\frac{1}{\sqrt{3}}(\ket{0000}+\ket{1111}+\ket{2222})$.}
\label{fig:fig3GHZ3d}
\end{figure}

As we have familiarized ourselves with the connection between graphs and quantum experiments \cite{krenn2017quantum}, we can use it to create higher-dimensional GHZ states, such as a 3-dimensional 4-particle GHZ state $\ket{GHZ_{4}^{3}}$. The corresponding graph is described in Fig. \ref{fig:fig3GHZ3d}A.

It has been shown in \cite{bogdanov267013, krenn2017quantum} that such a graph is the only graph which can be constructed where all perfect matchings are independent\footnote{Independent perfect matchings (which are called disjoint perfect matchings in graph theory), means that every edge appears exactly once in a perfect matching. If the perfect matchings contain common edges, we call them nonindependent perfect matchings.}. That means the quantum state $\ket{GHZ_{4}^{3}}$ is the only high-dimensional GHZ state which can be experimentally implemented in this way, while one can produce arbitrary 2-dimensional n-particle GHZ states $\ket{GHZ_{n}^{2}}$.

\begin{figure*}[!ht]
\centering
\includegraphics[width=\textwidth]{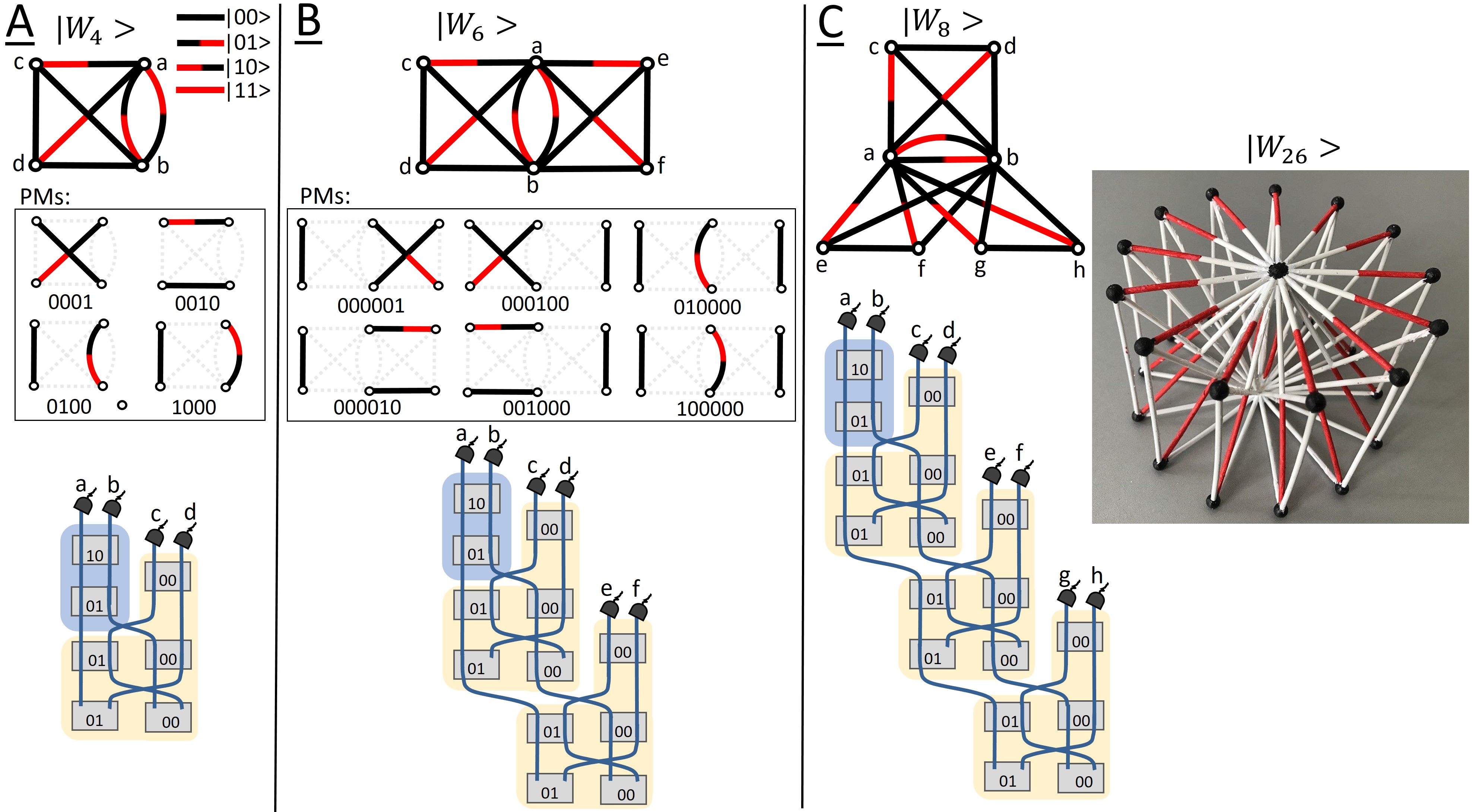}
\caption{General graphs and experiments for producing $n$-particle W states $\ket{W_{n}}$. \textbf{A}: A colored multigraph with four perfect matchings. Every perfect matching contains only one half-red [half dark gray] (black-red [black-dark gray] or red-black [dark gray-black]) edge, meaning that every term in the quantum state has exactly one excitation. The coherent superposition of all perfect matchings leads to a 4-particle W state $\ket{W_{4}}$. The corresponding experimental setup is described below the graph. \textbf{B} and \textbf{C}: In an analogous way, we show the graphs and experiments for generating 6- and 8-particle W states. On the right side, we show a 3D printed graph for a 26-particle W state $\ket{W_{26}}$. There the mode numbers $0$ and $1$ are represented with white and red [dark gray] colors, and the vertices are depicted in black. We call the graph for producing an $n$-photon W state $Oliver_n$ graph.}
\label{fig:fig4WstateDicke}
\end{figure*}

\section*{Generation of Dicke states}
One very large important class of states has been introduced by Robert H. Dicke, -- Dicke states $\ket{D^{k}_{n}}$ \cite{dicke1954coherence}. The states are defined as

\begin{equation}
\ket{D^{k}_{n}}=\frac{1}{\sqrt{\binom{n}{k}}}\hat{S}(\ket{0}^{\bigotimes(n-k)}\ket{1}^{\bigotimes k})
\end{equation}
where $n$ and $k$ stand for the number of particles and excitations, respectively. $\hat{S}$ is the symmetrical operator that gives summation over all distinct permutations of the $n$ particles.

\textit{W states $\ket{W_{n}}$} -- The special case with only one excitation is the well-known $n$-particle W state (denoted as $\ket{D^{1}_{n}}$ or $\ket{W_{n}}$) \cite{zeilinger1992higher, bourennane2004experimental}, which is highly persistent against photon loss. It is interesting that W states cannot be transformed into GHZ states with local operation and classical communication (LOCC) \cite{dur2000three}, meaning that they reside in different classes of entangled states.

Firstly we start with a 4-particle W state $\ket{W_{4}}$, which is
\begin{equation}
\ket{\psi}_{abcd}=\frac{1}{2}(\ket{0001}+\ket{0010}+\ket{0100}+\ket{1000}).
\label{eq:Wstate}
\end{equation}

\begin{figure*}[!t]
\centering
\includegraphics[width=0.78\textwidth]{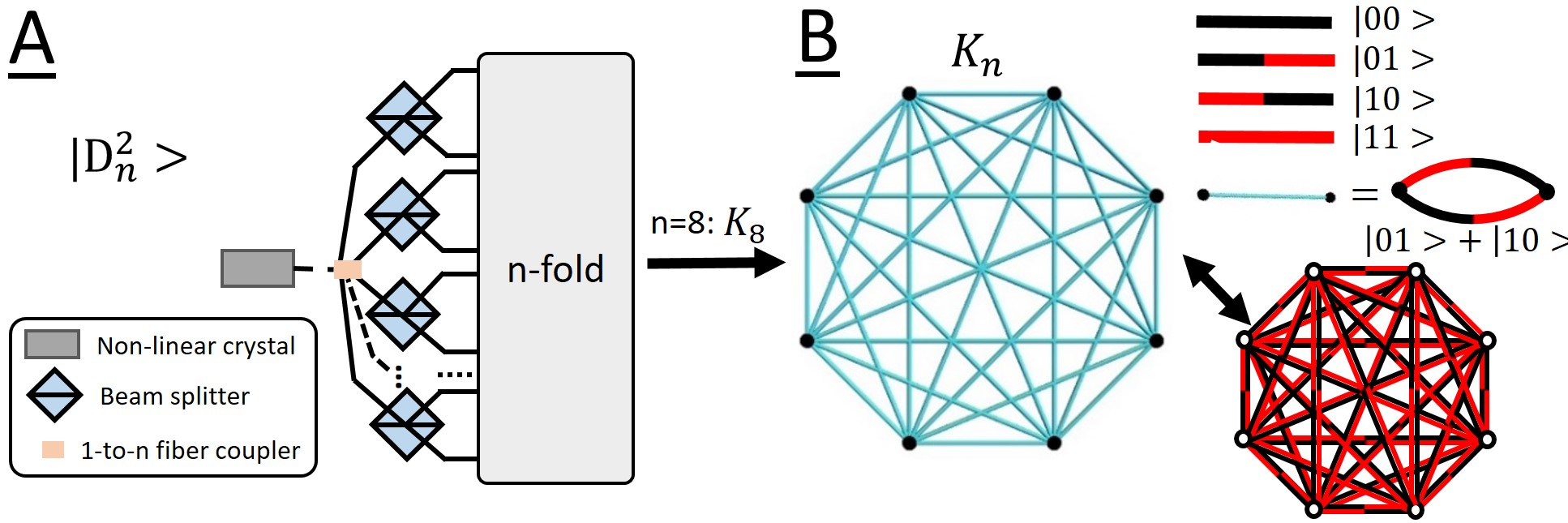}
\caption{General experimental scheme and graph for producing a symmetric Dicke state $\ket{D^{n/2}_{n}}$. \textbf{A}: One can pump a nonlinear crystal to create $n/2$ photon pairs. The photons of a photon pair have orthogonal mode numbers (such as horizontal and vertical polarization), which are denoted as mode numbers $0$ and $1$. These photon pairs are probabilistically separated by beam splitters. The Dicke state $\ket{D^{n/2}_{n}}$ is created conditionally on a click in every detector. \textbf{B}: In the general graph $K_{n}$, every blue [light gray] edge stands for a double edge with coloring black-red [black-dark gray] and red-black [dark gray-black], which corresponds to the state $\ket{01}+\ket{10}$. There, every perfect matching contains exactly $n/2$ half-red [half dark gray] (black-red [black-dark gray] or red-black [dark gray-black]) edges, which means each term in the quantum state includes $n/2$ excitations. Thus the superposition of all perfect matchings describes the Dicke state $\ket{D^{n/2}_{n}}$.}
\label{fig:fig5symmetryDicke}
\end{figure*}

\begin{figure*}[!t]
\centering
\includegraphics[width=0.78\textwidth]{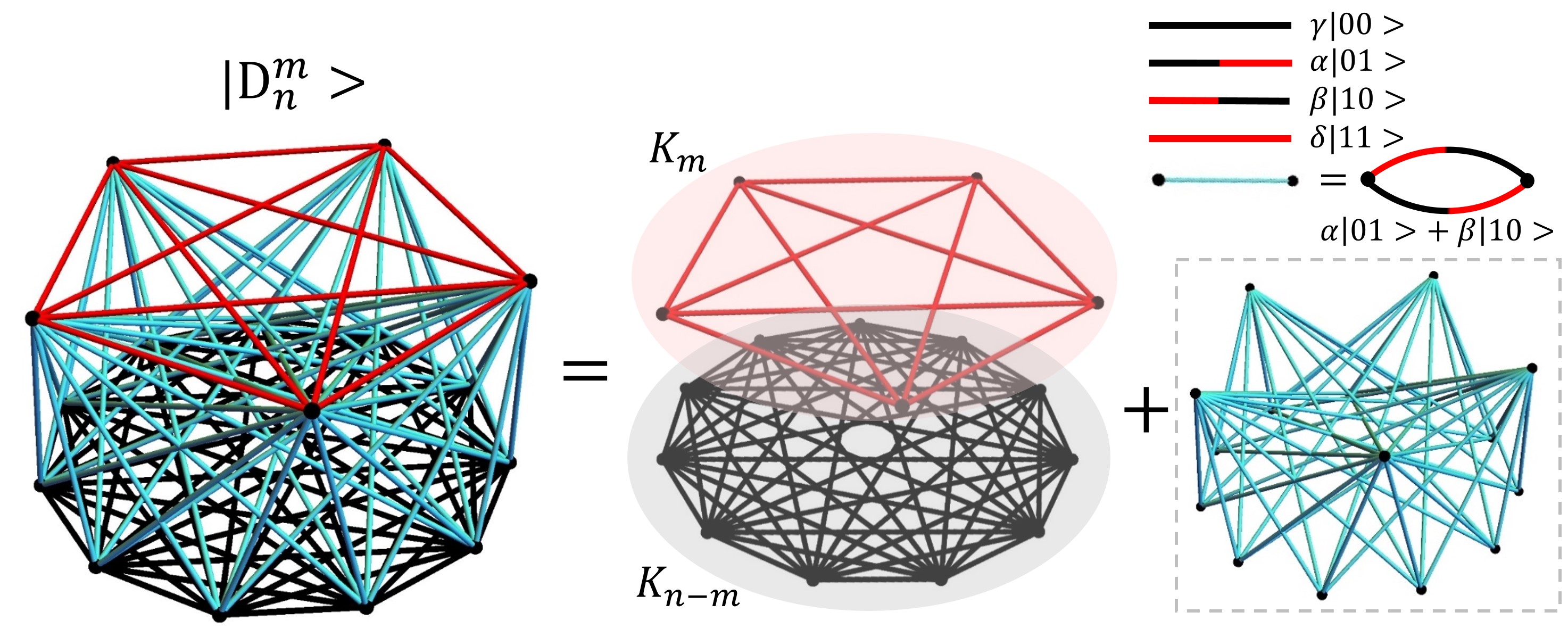}
\caption{General graph for general Dicke states $\ket{D^{m}_{n}}$ ($0 < m < n$). For better visualization, we show such a graph in a 3D viewpoint. The graph consists of two complete graphs $K_{n-m}$ and $K_{m}$. The first graph $K_{n-m}$ contains black edges while the second $K_{m}$ (upper) involves red [dark gray] edges. Each vertex of $K_{m}$ is connected to every vertex of $K_{n-m}$ with a blue [light gray] edge, which stands for a double edge consisting of a black-red [black-dark gray] edge and a red-black [dark gray-black] edge. We call a black-red [black-dark gray] edge or red-black [dark gray-black] edge as a half-red [half dark gray] edge. Thus a red [dark gray] edge (or red-red [dark gray-dark gray] edge) represents for two half-red [half dark gray] edges. Because of this construction, every perfect matching contains exactly $m$ half-red [half dark gray] edges, meaning that each term in the quantum state has $m$ excitations.}
\label{fig:fig6arbitraryDicke}
\end{figure*}

There are four terms in the quantum state, which correspond to four perfect matchings in the graph. For a complete graph\footnote{If every pair of vertices is connected with edges exactly once in a graph, we call such a graph as a complete graph. A complete graph with $n$ vertices is denoted as $K_{n}$.} $K_{4}$, the number of perfect matchings is three. However, we can use multiple edges to increase the number of perfect matchings. These graphs are denoted as multigraphs.

We show such a multigraph for the W state $\ket{W_{4}}$ in Fig. \ref{fig:fig4WstateDicke}A. There, every edge can contain two colors (black and red [dark gray]). For example, a red [dark gray] edge stands for that the corresponding crystal produces photon pairs in a state $\ket{11}$. Thus the edges with colors black-black, black-red [black-dark gray], red-black [dark gray-black] and red-red [dark gray-dark gray] represent the states $\ket{00}$, $\ket{01}$, $\ket{10}$ and $\ket{11}$, respectively.

We find that every perfect matching contains only one half-red [half dark gray] (black-red [black-dark gray] or red-black [dark gray-black]) edge and no more other red [dark gray] edges can be involved. That means every term in the quantum state contains exactly one excitation and their coherent superposition describes a W state. The corresponding optical setup is described below the graph. Therefore, one can experimentally produce 4-particle W state $\ket{W_{4}}$ \cite{krenn2017quantum}.

Now we generalize the graph for arbitrary $n$-particle W state $\ket{W_{n}}$. We connect all the half-red [half dark gray] edges to vertex $a$ and describe the graphs in Fig. \ref{fig:fig4WstateDicke}. Thereby, every perfect matching contains exactly one half-red [half dark gray] edge because of the fact that vertex $a$ can be used only once in a perfect matching. This gives exactly one excitation in every term of the quantum state. Thus one can construct such graphs for producing arbitrary W states. A 3D printed graph for a 26-particle W state $\ket{W_{26}}$ is shown in Fig. \ref{fig:fig4WstateDicke}C.

Interestingly, the structure of the graph for creating $n$-particle W state $\ket{W_{n}}$ can be seen as a strong product of graphs \cite{sabidussi1959graph, weissteingraphproduct}. The general graph for state $\ket{W_{n}}$ is a special book graph \cite{weissteinbookgraph}, which consists of $n/2-1$ complete graphs $K_{4}$ with common edges $E_{ab}$ (for details see the Appendix A). The multiple common edge $E_{ab}$ is the so-called base of the book graph and the $n/2-1$ complete graphs form the pages of our book graph. Hence such a graph can also be called a \textit{(n/2-1)-page 2-base $K_{4}$-book graph} \cite{drorgraphname294174}. For simplicity, we denote such a graph as an \textit{Oliver$_n$} graph. Thus, the graph for W state $\ket{W_{8}}$, which is shown in Fig. \ref{fig:fig4WstateDicke}C, is a book graph with three pages.

\textit{Dicke states $\ket{D^{n/2}_{n}}$} -- Another special case of Dicke states, which has been experimentally investigated, are the states $\ket{D^{n/2}_{n}}$. By splitting probabilistically photons, experimental implementations for Dicke states $\ket{D^{2}_{4}}$ and $\ket{D^{3}_{6}}$ have been successfully realized in laboratories \cite{kiesel2007experimental, prevedel2009experimental, wieczorek2009experimental, hiesmayr2016observation}. The general experimental scheme for symmetric Dicke states $\ket{D^{n/2}_{n}}$ is described in Fig. \ref{fig:fig5symmetryDicke}A.
\begin{figure}[!t]
\centering
\includegraphics[width=0.46\textwidth]{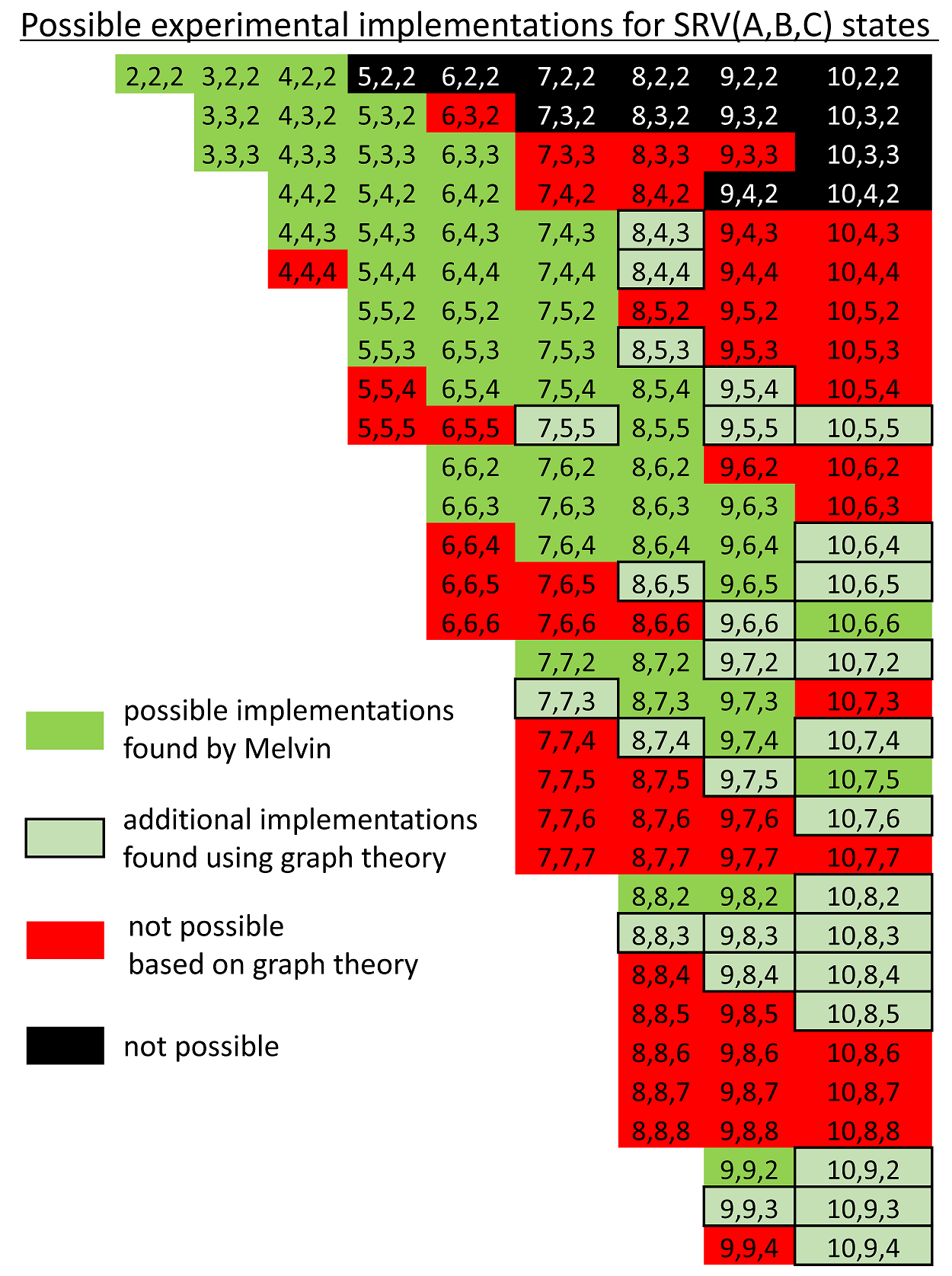}
\caption{A list of experimentally possible $SRV(A, B, C)$ states. Strong green [light gray] cells show that these states have been found with the computer algorithm \melvin \cite{krenn2016automated}. For all remaining cases, using graph theory we find the corresponding experimental setups (light green [light gray in solid box]) or that states cannot be created (red [dark gray]) with probabilistic sources (without additional ancillary photons). States represented by black cells cannot exist even theoretically, because of combinatorial reasons \cite{huber2013structure}.}
\label{fig:fig7SRVTable}
\end{figure}

The corresponding graph for such experimental setup is a complete graph $K_{n}$, which is described in Fig. \ref{fig:fig5symmetryDicke}B. There every pair of vertices is connected with a blue [light gray] edge, which stands for multiple edges colored with black-red [black-dark gray] and red-black [dark gray-black]. Therefore, every perfect matching contains $n/2$ half-red [half dark gray] edges, meaning that each term in the quantum state involves $n/2$ excitations. The coherent superposition of all perfect matchings describes the symmetric Dicke state $\ket{D^{n/2}_{n}}$.

\textit{General Dicke states $\ket{D^{m}_{n}}$} -- A natural question is whether we can experimentally create arbitrary Dicke states $\ket{D^{m}_{n}}$ ($0<m<n$). We answer the question affirmatively, and show the construction of a graph in Fig. \ref{fig:fig6arbitraryDicke}. In general, we use two complete graphs $G_{1}=K_{n-m}$ and $G_{2}=K_{m}$, where all edges of $G_{1}$ are black while all edges of $G_{2}$ are red [dark gray]. Each vertex of $G_{1}$ is connected to every vertex of $G_{2}$ with a blue [light gray] edge, which is a double edge with red-black [dark gray-black] and black-red [black-dark gray].

While in all constructions before, all terms of the resulting quantum state had the same amplitude (which we call maximally entangled), that is not the case here anymore. In quantum experiments, one can adjust the pump power to make nonmaximally entangled states into maximally entangled states, which means adjusting all amplitudes to be the same values. For such quantum state, the total number of terms in the quantum state is given by the number of perfect matchings of the corresponding graphs, which holds for the rest of the paper. These introduce weights in the graphs, which have been investigated in \cite{gu2019quantum}. We show some examples of maximally entangled Dicke states in the Appendix B.

\section*{generation of high-dimensional multipartite entangled states}
The generalization of high-dimensional entangled states allows very rich types of nonclassical correlations. One method to characterize these states is the so-called Schmidt-Rank Vector (SRV) \cite{huber2013structure, huber2013entropy, cadney2014inequalities}. These states give rise to asymmetrically entangled states that exist only if both the number of particles and the dimensions are larger than two. We study one important special case of 3-particle entangled states with an additional particle as a trigger. These states recently have been investigated experimentally \cite{malik2016multi, erhard2018experimental}, and studied extensively in the form of computer-designed experiments \cite{krenn2016automated, melnikov2018active}.

The SRV represents the rank of the reduced density matrices of each particle. In the quantum state of three parties $a$, $b$ and $c$, the rank of the reduced density matrices
\begin{align*}
A=rank(Tr_{a}(\ket{\psi}\bra{\psi}))\\
B=rank(Tr_{b}(\ket{\psi}\bra{\psi}))\\
C=rank(Tr_{c}(\ket{\psi}\bra{\psi}))
\label{eq:SRVrank}
\end{align*}
together form the SRV $d_{\psi}=(A, B, C)$, where $A\geq B\geq C$. The values $A$, $B$ and $C$ stand for the dimensionality of entanglement particle $a$, $b$ and $c$ with the other two parties.

The classification with different SRVs provides an interesting insight that one can transform quantum states from higher classes to lower classes with LOCC, and not vice versa\footnote{The dimensionality $i$ ($i=A, B, C$) cannot be increased with LOCC.}.

As an example, we show a maximally entangled state with SRV=(4, 2, 2), which is
\begin{equation}
\ket{\psi}_{abc}=\frac{1}{2}(\ket{000}+\ket{101}+\ket{210}+\ket{311}).
\label{eq:SRVexample}
\end{equation}
There the first particle $a$ is 4-dimensionally entangled with the other two particles $bc$, whereas particle $b$ and $c$ are both only 2-dimensionally entangled with the rest.

We are interested in maximally entangled states (as before, all amplitudes are the same). Furthermore, we want that the quantum state with $SRV(A, B, C)$ has $A$ terms. Thereby, the structure of the SRV is clearly visible in the computation basis, which is convenient experimentally. We call such an entangled state an $SRV(A, B, C)$ state.

Searching experimental implementations for producing $SRV(A, B, C)$ states has been investigated with the computer algorithm \melvin \cite{krenn2016automated}. In Fig. \ref{fig:fig7SRVTable}, for the strong green cells, \melvin has found experimental setups after several months of runtime. All other cases have remained open.

Now one could ask which $SRV(A, B, C)$ states are experimentally possible to create with probabilistic photon pair sources? We apply our connection between graphs and quantum experiments to answer the question. In \cite{krenn2017quantum}, the authors have shown that graphs with four vertices can contain maximally three independent perfect matchings. We extend that technique and find whether one can experimentally create an $SRV(A, B, C)$ state without additional particles with probabilistic pair sources (details see the Appendix C)\footnote{All of the experimental setups are based on Ref. \cite{krenn2017entanglement}. It is an open question how to create these setups with nonlinear crystals producing photon pairs and linear optics.}. This finally answers a question that has been open for 3 years.

Our technique can be applied to find experimental implementations for another type of high-dimensional multipartite quantum states such as absolutely maximally entangled state \cite{scott2004multipartite, goyeneche2014genuinely, goyeneche2015absolutely, fhuberAME35}. We show more interesting examples in the Appendix D. Many related questions remain open, and are summarized elsewhere \cite{krenn2019questions}.

\section*{Conclusion}

We have presented a method to experimentally create large classes of entangled quantum states that are theoretically well studied but unexplored in laboratories, by extending recent ideas in Ref. \cite{krenn2017entanglement} and the bridge between quantum experiments and graphs \cite{krenn2017quantum}.

An exciting extension of our work would be a full classification of which quantum states are achievable with current photonic technology involving probabilistic pair sources.

One particular important class of photonic entangled states are so-called \textit{graph states}, which are resources for measurement-based quantum computation \cite{raussendorf2001one, raussendorf2003measurement}. Despite the similarity of names, \textit{graph states} are not related to the techniques explained here. It would be very interesting to investigate which type of \textit{graph states} can be experimentally generated with probabilistic pair sources. A starting point will be the introduction of complex weights, which has been discussed in Ref. \cite{gu2019quantum}.

Motivated by our results, another purely physical question raises: What does it mean physically that some entangled quantum states cannot be created? Is the producibility or lack thereof connected to a property of entanglement, such as entanglement of formation \cite{wootters1998entanglement}? While the graph theoretical representation covers the mathematical results in an excellent way, a physical interpretation of these results is still missing. It would be an exciting research project to shed more light on that question.

\section*{Acknowledgements}
This work was supported by the Austrian Academy of Sciences (\"OAW), by the Austrian Science Fund (FWF) with SFB F40 (FOQUS), the National Natural Science Foundation of China (No.61771236) and its Major Program (No. 11690030, 11690032), the National Key Research and Development Program of China (2017YFA0303700), and from a Scholarship from the China Scholarship Council (CSC).

\appendix

\section{Strong Product of Graphs}

\begin{figure}[!t]
\centering
\includegraphics[width=0.47\textwidth]{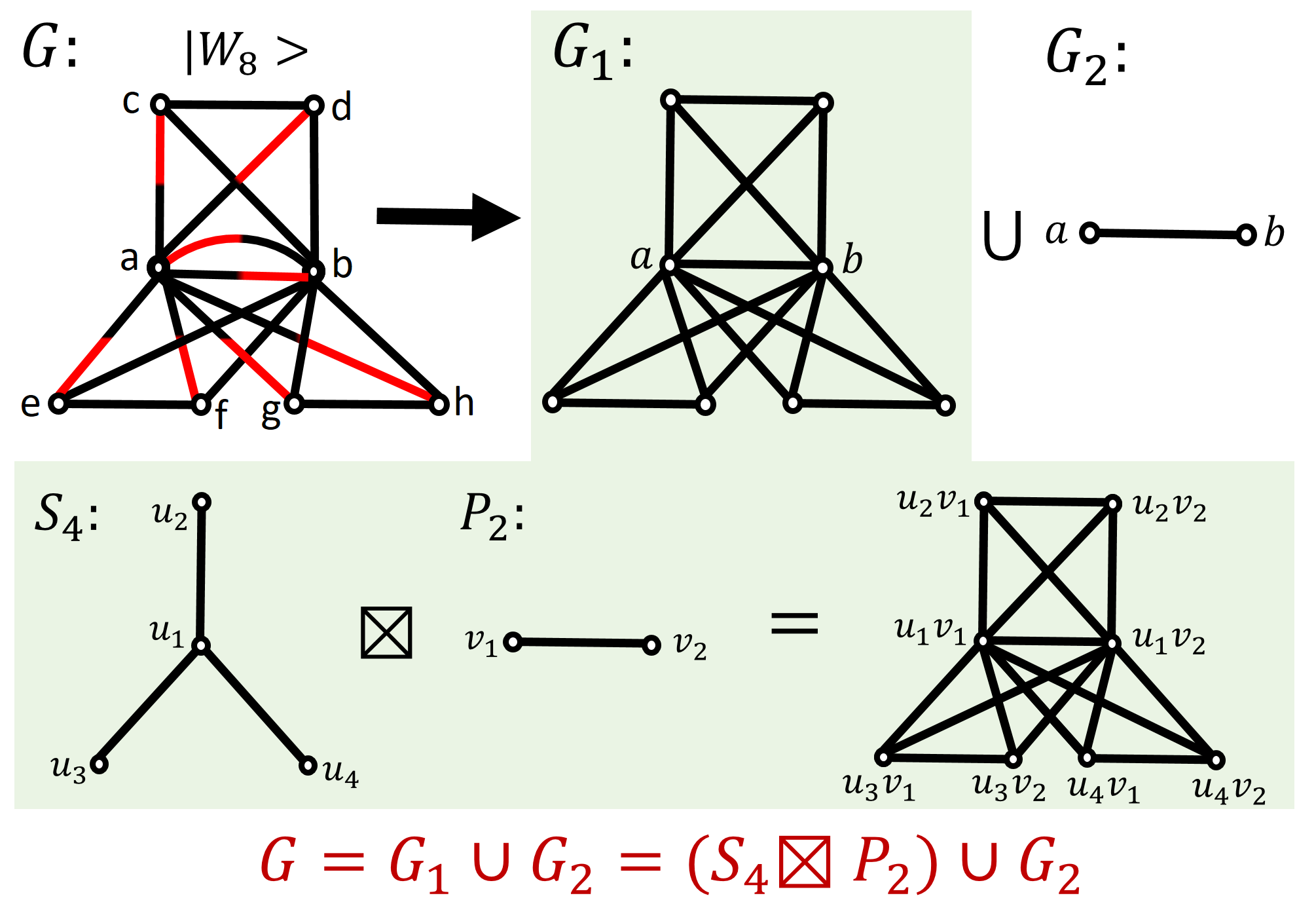}
\caption{Graph operations for constructing the graph corresponding to an $n$-particle W state -- the $Oliver_n$ graph. The structure of graph $G$ for the W state $\ket{W_{8}}$ can be seen as a union of graphs $G_{1}$ and $G_{2}$, which is $G=G_{1}\bigcup G_{2}$. Graph $G_{1}$ is a strong product of a star graph $S_{4}$ and a path graph $P_{2}$, which is $G_{1}=S_{4} \boxplus P_{2}$. Therefore, the graph for $n$-particle W state $\ket{W_{n}}$ can be depicted as $G = (S_{i} \boxplus P_{2}) \bigcup G_{2}$, where $i=n/2$.}
\label{fig:figS1bookgraph}
\end{figure}

Here we explain the structure of graphs for $n$-particle W states $\ket{W_{n}}$ and show a graph $G$ for the W state $\ket{W_{8}}$ in Fig. \ref{fig:figS1bookgraph}.

The graph $G$ can be seen as the result of a union operation of graphs $G_{1}$ and $G_{2}$. There the graph $G_{1}$ is a strong product\footnote{The strong product $G \boxplus H$ of graphs $G$ and $H$ is the graph with vertex set $V(G)\times V(H)$ and u=(u1,v1) is adjacent with v=(u2,v2) whenever (v1=v2 and u1 is adjacent with u2) or (u1=u2 and v1 is adjacent with v2) or (u1 is adjacent with u2 and v1 is adjacent with v2).} \cite{sabidussi1959graph, weissteingraphproduct} ($S_{4} \boxplus P_{2}$) of a star graph\footnote{A star graph $S_{i}$ is a graph with $i$ vertices, where ($i-1$) vertices are only connected, with one edge, to a single central vertex.} $S_{i}$ and a path graph\footnote{A path graph $P_{j}$ is a graph with $j$ vertices, where $j$ vertices and $(j-1)$ edges lie on a single line.} $P_{i}$. The graph $G_{1}$ can be seen as a special case of book graph \cite{weissteinbookgraph}. The graph $P_{2}$ is the base of the book graph and the number of edges of the graph $S_{i}$ gives the number of pages in the book graph. Therefore, the graph for the quantum state $\ket{W_{8}}$ is a book graph with three pages.

\begin{figure*}[!t]
\centering
\includegraphics[width=\textwidth]{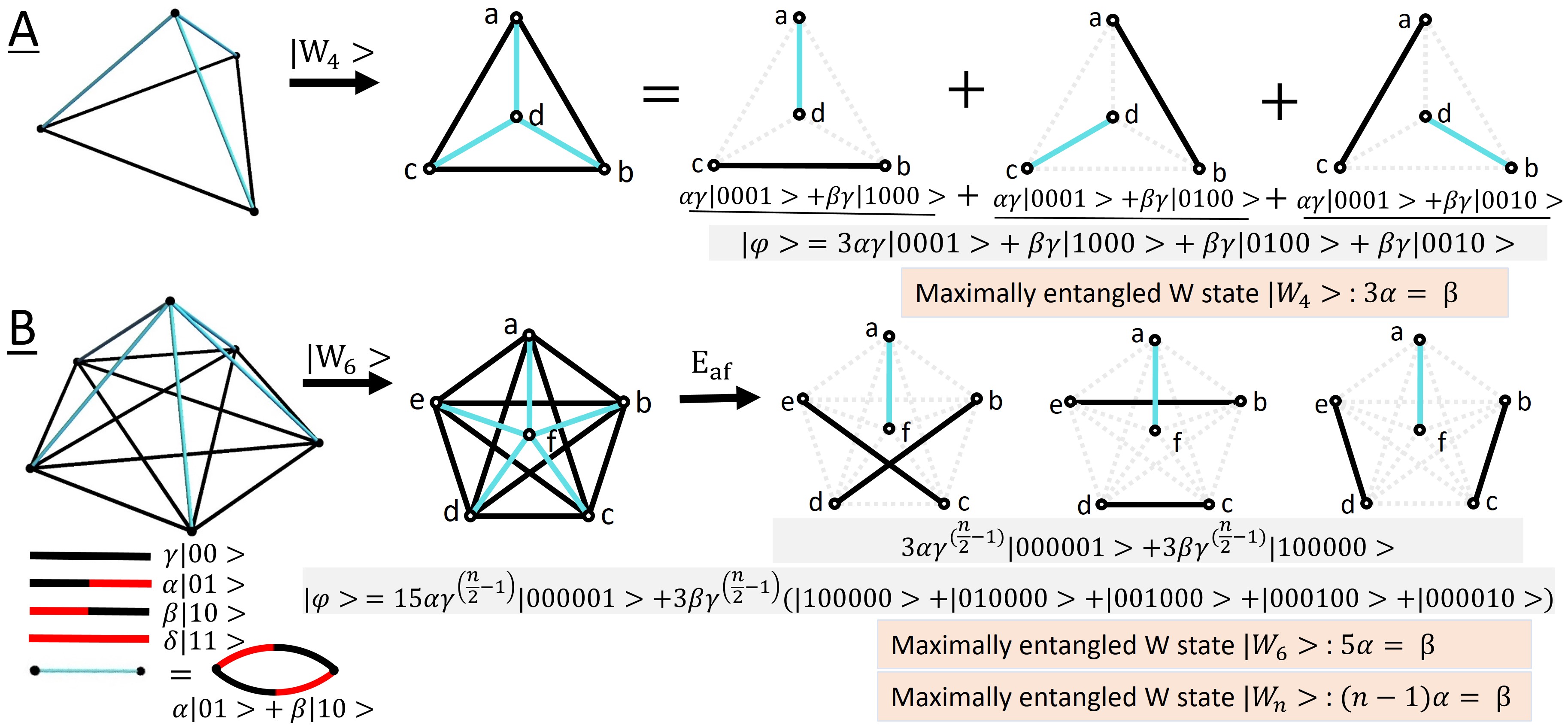}
\caption{Weighted graphs for $n$-particle W states $\ket{W_{n}}$. \textbf{A}: Graph for W state $\ket{W_{4}}$. The weights for black-black, black-red [black-dark gray], red-black [dark gray-black] and red-red [dark gray-dark gray] edges are $\delta$, $\alpha$, $\beta$ and $\gamma$, respectively. For simplicity, we write the black-red [black-dark gray] and red-black [dark gray-black] edges as a blue [light gray] edge, which corresponds to a state $\alpha\ket{01}+\beta\ket{10}$. From Fig. \ref{fig:fig6arbitraryDicke}, we know that such a graph can be redrawn as two complete graphs $K_{1}$ and $K_{3}$ connected with blue [light gray] edges. The coherent superposition of all the perfect matchings in such a graph leads to the final quantum state, which is $\ket{\psi}_{abcd}=\gamma(3\alpha\ket{0001}+\beta\ket{0010}+\beta\ket{0100}+\beta\ket{1000})$ (without normalization). In order to obtain the maximally entangled W state, the coefficients should be the same, meaning $3\alpha=\beta$. For example, we can set the weights ($\alpha=1, \beta=3, \gamma=1$). \textbf{B}: Graph for W state $\ket{W_{6}}$. In an analogous way, we need to calculate all the perfect matchings of the graph. Firstly we start with edge $E_{af}$. The graph can be decomposed into the edge $E_{af}$ and a graph $K_{n-2}$ with vertices $b$, $c$, $d$ and $e$. In this case, the superposition of the perfect matchings are calculated, which is $3\gamma^{2}(\alpha\ket{000001}+\beta\ket{100000})$. We calculate all the perfect matchings and require that $5\alpha=\beta$. Then we obtain the W state. In general, we can obtain $n$-particle W states with $(n-1)\alpha=\beta$.}
\label{fig:figS3weightedGraph1}
\end{figure*}

\begin{figure*}[!t]
\centering
\includegraphics[width=\textwidth]{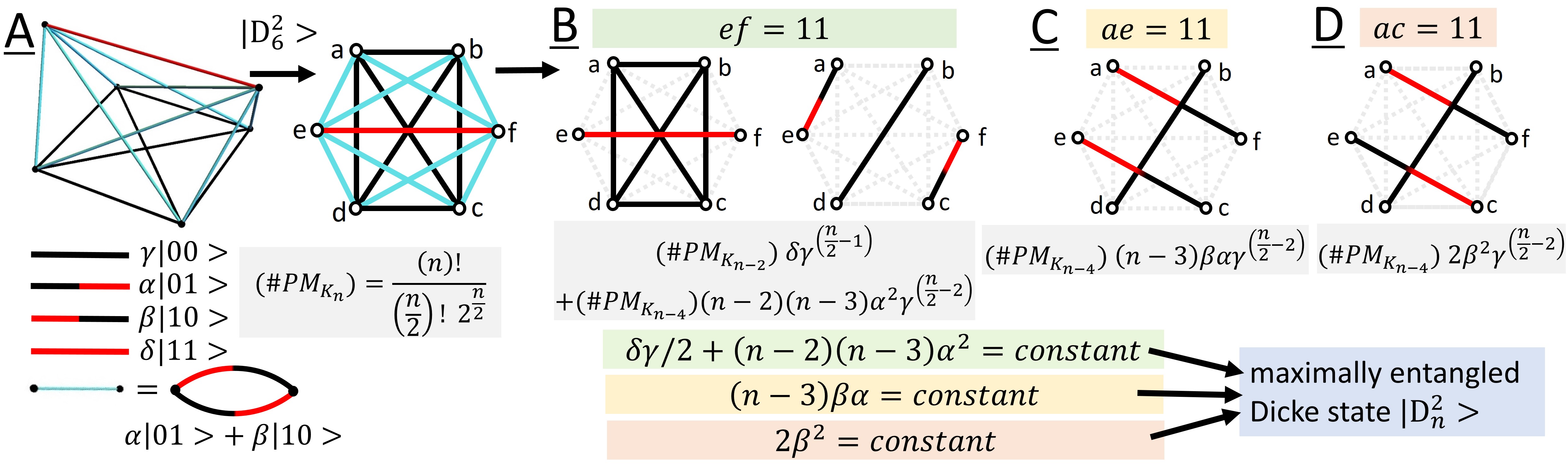}
\caption{Weighted graph for Dicke states $\ket{D^{2}_{n}}$. \textbf{A}: Graph for Dicke state $\ket{D^{2}_{6}}$. This graph can be redrawn as two complete graphs $K_{4}$ and $K_{2}$ connected with blue [light gray] edges. \textbf{B}: Firstly we consider the term $\ket{000011}$ in the Dicke state, we find that there are two cases in the graph where the perfect matchings lead to that term. One is that the perfect matchings contains red [dark gray] edge $E_{ef}$. The other case is that the red-black [dark gray-black] edges connect to vertices $e$ and $f$. Thereby, we find the number of perfect matchings leading to the term of the quantum state. \textbf{C}: Next we consider the terms where one of the vertices $e$ and $f$ and one of vertices ($a$, $b$, $c$ and $d$) are related to red [dark gray] coloring edge. Then, we enumerated the cases where two of vertices($a$, $b$, $c$ and $d$) are connecting to red [dark gray] coloring edges. Finally, by solving these equation systems, one can create Dicke states $\ket{D^{2}_{n}}$.}
\label{fig:figS4weightedGraph2}
\end{figure*}

\begin{figure*}[!t]
\centering
\includegraphics[width=\textwidth]{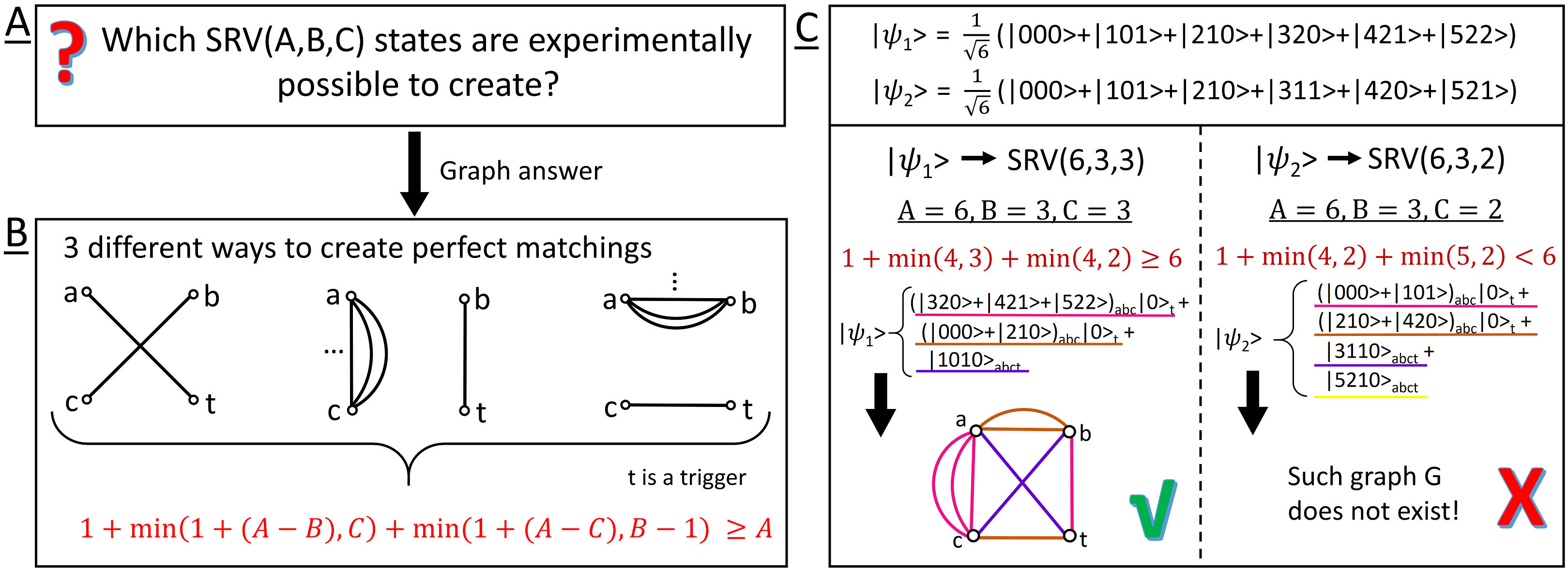}
\caption{Restriction on the generation of 3-particle maximally entangled states with different Schmidt-Rank Vectors ($SRV$) based on graph theory. \textbf{A}: We ask which $SRV(A, B, C)$ states can be created. \textbf{B}: Each term of an $SRV(A, B, C)$ state is given by a perfect matching of the corresponding graph. If a graph with four vertices involving more than $A$ perfect matchings can be constructed, a possible experimental setup for such a state exists. Therefore, when the parameters $A$, $B$ and $C$ fulfill the condition $1+min(1+(A-B),C)+min(1+(A-C),B-1)\geq A$, one could experimentally produce an $SRV(A, B, C)$ state. \textbf{C}: Two examples with different SRVs. In the case of $SRV(6, 3, 3)$, we apply the restriction and the parameters fulfill the condition derived in \textbf{B}, thus such a state can be created. However, in the case of $SRV(6, 3, 2)$, the parameters do not fulfill the requirement and one needs four disjoint perfect matchings of a graph with four vertices. Such a graph does not exist. Therefore, this quantum state cannot be produced with probabilistic photon pair sources in such way.}
\label{fig:figS5SRVProof}
\end{figure*}

\section{Graphs for General Dicke States}
We have shown a general graph for arbitrary nonmaximally Dicke states in Fig. \ref{fig:fig6arbitraryDicke}. Each term in the quantum state corresponds to a number of perfect matchings. The number of perfect matchings is not necessarily the same number of the terms in the quantum state. Experimentally this leads to different coefficients for each term of the state and thereby to nonmaximally entanglement.

In the laboratories, one can adjust the pump power to change the amplitudes in order to obtain the maximally entangled states. This will introduce weights in the corresponding graph \cite{gu2019quantum}. We show how to make the nonmaximally Dicke states $\ket{D^{1}_{n}}$ and $\ket{D^{2}_{n}}$ to the maximally entangled Dicke states in Figs. \ref{fig:figS3weightedGraph1} and \ref{fig:figS4weightedGraph2}.

We do this by computing all perfect matchings that correspond to individual terms and then require that the corresponding weights lead to a constant value. This leads to an algebraic equation system. For the example mentioned above, that system can be solved.

\section{Restriction on the Generation of $\mathbf{SRV(A, B, C)}$ States}

Here we apply the connection between graphs and experiments to answer which maximally entangled $SRV(A, B, C)$ states can be created. As we have described in the main text, for an $SRV(A, B, C)$ state with an additional trigger $t$ ($t$ stays the same mode number), the dimensionality of particles $a$, $b$ and $c$ are given by the values $A$, $B$ and $C$. That means particles $a$, $b$ and $c$ must contain $A$, $B$ and $C$ different mode numbers ($A\geq B\geq C$).

In the graph description, every perfect matching of the graph corresponds to a term in the quantum state. Thus we need to construct a graph with exactly $A$ perfect matchings, as this is part of our definition of maximally entanglement. We now use the three disjoint perfect matchings that exist in the complete graph $K_4$ to find a possible experimental implementation for different $SRV(A, B, C)$ states.

The main idea is, when there is more than one term with the same mode number for a particle $a$, $b$ or $c$, we could combine the trigger $t$ together with the particle of the repeated mode number to form a multiedge. That will allow us to create more than three terms in the quantum state (Note: we can always create three arbitrary terms, as we have full control of edges in the three perfect matchings.). In total we need to create $A$ terms.

First we consider the edge $E_{t,a}$. The mode number in each term of particle $a$ needs to be different, thus we can only use $E_{t,a}$ to create one term.

Now we consider the edge $E_{t,b}$. Photon $b$ has $B$ different mode numbers, therefore in $A-B$ terms, the mode numbers can be the same. So in addition to the one term that we always create, we have the possibility to create $A-B$ additional terms, leading to $1+(A-B)$ terms producible using $E_{t,b}$. However, in the cases when we use the same mode number for particle $b$, the mode number for particle $c$ needs to be different (otherwise it would reduce the dimensionality of the state, for example: $\ket{0}_a \ket{0,0}_{b,c}$ + $\ket{1}_a \ket{0,0}_{b,c}$ = $\left(\ket{0}_a + \ket{1}_a \right) \ket{0,0}_{b,c}$. That means there is a tradeoff between the number of repetitions in particle $b$ that we can use, and the number of different modes particle $c$ has (which is $C$). So in total, using edge $E_{t,b}$, we can create $\min(1+(A-B),C)$ terms.

Finally, we apply the same argument to the terms that we can create using $E_{t,c}$. We use the $(A-C)$ repetitions to create $1+(A-C)$ terms, again conditioned that there are enough usable mode numbers of photon $b$. That usable numbers of different modes in $b$ is now $(B-1)$, because one mode number was already used in the perfect matchings using $E_{t,b}$. Therefore we find that, using the edge $E_{t,c}$, we can create $\min(1+(A-C),B-1)$ terms.

Overall we find the following condition explaining whether the $SRV(A,B,C)$ can be created:
\begin{equation}
1+min(1+(A-B),C)+min(1+(A-C),B-1) \geq A
\label{eq:SRVconditionSupp}
\end{equation}

We illustrate that conditions in Fig. \ref{fig:figS5SRVProof} and describe two concrete examples in Fig. \ref{fig:figS5SRVProof}C.

\begin{figure}[!t]
\centering
\includegraphics[width=0.45\textwidth]{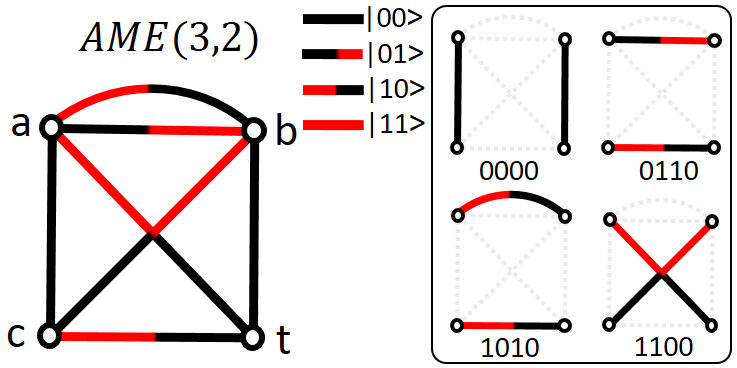}
\caption{A graph for a possible experiment producing the $AME(3,2)$ state with an additional trigger. There are four perfect matchings in the graph, which are described in the solid box. The coherent superposition of all perfect matchings leads to the quantum state $\ket{\psi}_{abct}=\frac{1}{2}(\ket{000}+\ket{011}+\ket{101}+\ket{110})\ket{0}$.}
\label{fig:figS6AMEstate}
\end{figure}

\section{Generation of Absolutely Maximally Entangled States}

The absolutely maximally entangled (AME) states are another type of multipartite states, which give the maximally mixed states by tracing out half or more of the parties. Such state is defined as $AME(n,d)$ with $n$ particles of local dimension $d$.

Here we only consider the experimentally most significant cases with $n=3$, which is written as \cite{goyeneche2018entanglement}
\begin{equation}
\ket{AME(3,d)}=\frac{1}{d}\sum_{\substack{i=0\\j=0}}^{d-1} \ket{i,j,i+j}
\label{eq:AMEstate}
\end{equation}
where sums inside kets are computed to be modulo $d$.

Firstly, we consider the 2-dimensional 3-particle AME state, which is
\begin{equation}
\ket{\psi}_{abc}=\frac{1}{2}(\ket{000}+\ket{011}+\ket{101}+\ket{110}).
\label{eq:AME32state}
\end{equation}

Here, we apply the technique from the restriction for creating $SRV(A, B, C)$ states in Fig. \ref{fig:figS5SRVProof}. Thus we can rewrite such a state in Eq. \ref{eq:AME32state} as

\begin{align}
\ket{\psi}_{abct}=\frac{1}{2}&(\ket{0000}+\ket{0110}+\ket{1010}+\ket{1100})\notag\\
=\frac{1}{2}&\{\left(\ket{01}+\ket{10}\right)_{ab}\ket{10}_{ct}+\ket{0000}_{abct}+\ket{1100}_{abct}\}.
\end{align}

We show such a graph in Fig. \ref{fig:figS6AMEstate}, which means that the quantum state can be experimentally produced.

Now we consider a 3-dimensional 3-particle AME state, which is
\begin{align}
\ket{\psi}_{abc}=\frac{1}{3}&(\ket{000}+\ket{011}+\ket{022}+\ket{101}+\ket{112}\notag \\
+&\ket{120}+\ket{202}+\ket{210}+\ket{221}\notag).
\end{align}

In an analogous way, such a state can be rewritten as
\begin{align}
\ket{\psi}_{abct}=\frac{1}{3}&\{(\ket{11}+\ket{22})\ket{00}_{bt}+(\ket{12}+\ket{21})\ket{00}_{ct}\notag \\
+&(\ket{00}+\ket{11}+\ket{22})_{bc}\ket{00}_{at}+\ket{1120}_{abct}\notag \\
+&\ket{2210}_{abct}\}
\end{align}

There we would need more than three independent perfect matchings for a graph with 4-vertices. However such a graph does not exist. Thus one cannot experimentally produce the state $\ket{AME(3,3)}$ in such a way. Similarly, the state $\ket{AME(4,d)}$ with $d \geq 2$ cannot be created in this way.

\bibliographystyle{unsrt}
\bibliography{refs}

\begin{thebibliography}{}

\bibitem{einstein1935can}
A. Einstein, B. Podolsky and N. Rosen, Can quantum-mechanical description of
  physical reality be considered complete?. \textit{Physical review}
  \textbf{47}, 777 (1935).

\bibitem{bell1964einstein}
J.S. Bell, On the einstein podolsky rosen paradox. \textit{Physics} \textbf{1},
  195 (1964).

\bibitem{greenberger1989going}
D.M. Greenberger, M.A. Horne and A. Zeilinger, Going beyond Bell's theorem.
  \textit{Bells Theorem, Quantum Theory and Conceptions of the Universe,
  Fundamental Theories of Physics, edited by M. Kafatos (Springer, Dordrecht,
  1989)} \textbf{37}, 69--72 (1989).

\bibitem{greenberger1990bell}
D.M. Greenberger, M.A. Horne, A. Shimony and A. Zeilinger, Bell's theorem
  without inequalities. \textit{American Journal of Physics} \textbf{58},
  1131--1143 (1990).

\bibitem{dicke1954coherence}
R.H. Dicke, Coherence in spontaneous radiation processes. \textit{Physical
  Review} \textbf{93}, 99 (1954).

\bibitem{ryu2014multisetting}
J. Ryu, C. Lee, Z. Yin, R. Rahaman, D.G. Angelakis, J. Lee and M. {\.Z}ukowski,
  Multisetting Greenberger-Horne-Zeilinger theorem. \textit{Physical Review A}
  \textbf{89}, 024103 (2014).

\bibitem{lawrence2014rotational}
J. Lawrence, Rotational covariance and Greenberger-Horne-Zeilinger theorems for
  three or more particles of any dimension. \textit{Physical Review A}
  \textbf{89}, 012105 (2014).

\bibitem{erhard2018experimental}
M. Erhard, M. Malik, M. Krenn and A. Zeilinger, Experimental
  Greenberger--Horne--Zeilinger entanglement beyond qubits. \textit{Nature
  Photonics} \textbf{12}, 759-764 (2018).

\bibitem{huber2013structure}
M. Huber and J.I. Vicente, Structure of multidimensional entanglement in
  multipartite systems. \textit{Physical Review Letters} \textbf{110}, 030501
  (2013).

\bibitem{goyeneche2016multipartite}
D. Goyeneche, J. Bielawski and K. {\.Z}yczkowski, Multipartite entanglement in
  heterogeneous systems. \textit{Physical Review A} \textbf{94}, 012346 (2016).

\bibitem{malik2016multi}
M. Malik, M. Erhard, M. Huber, M. Krenn, R. Fickler and A. Zeilinger,
  Multi-photon entanglement in high dimensions. \textit{Nature Photonics}
  \textbf{10}, 248 (2016).

\bibitem{pivoluska2018layered}
M. Pivoluska, M. Huber and M. Malik, Layered quantum key distribution.
  \textit{Physical Review A} \textbf{97}, 032312 (2018).

\bibitem{krenn2017quantum}
M. Krenn, X. Gu and A. Zeilinger, Quantum experiments and graphs: Multiparty
  states as coherent superpositions of perfect matchings. \textit{Physical
  Review Letters} \textbf{119}, 240403 (2017).

\bibitem{krenn2016automated}
M. Krenn, M. Malik, R. Fickler, R. Lapkiewicz and A. Zeilinger, Automated
  search for new quantum experiments. \textit{Physical Review Letters}
  \textbf{116}, 090405 (2016).

\bibitem{melnikov2018active}
A.A. Melnikov, H.P. Nautrup, M. Krenn, V. Dunjko, M. Tiersch, A. Zeilinger and
  H.J. Briegel, Active learning machine learns to create new quantum
  experiments. \textit{Proceedings of the National Academy of Sciences}
  \textbf{115}, 1221--1226 (2018).

\bibitem{krenn2017entanglement}
M. Krenn, A. Hochrainer, M. Lahiri and A. Zeilinger, Entanglement by path
  identity. \textit{Physical Review Letters} \textbf{118}, 080401 (2017).

\bibitem{allen1992orbital}
L. Allen, M.W. Beijersbergen, R. Spreeuw and J. Woerdman, Orbital angular
  momentum of light and the transformation of Laguerre-Gaussian laser modes.
  \textit{Physical Review A} \textbf{45}, 8185 (1992).

\bibitem{yao2011orbital}
A.M. Yao and M.J. Padgett, Orbital angular momentum: origins, behavior and
  applications. \textit{Advances in Optics and Photonics} \textbf{3}, 161--204
  (2011).

\bibitem{krenn2017orbital}
M. Krenn, M. Malik, M. Erhard and A. Zeilinger, Orbital angular momentum of
  photons and the entanglement of Laguerre--Gaussian modes. \textit{Phil.
  Trans. R. Soc. A} \textbf{375}, 20150442 (2017).

\bibitem{franson1989bell}
J.D. Franson, Bell inequality for position and time. \textit{Physical Review
  Letters} \textbf{62}, 2205 (1989).

\bibitem{versteegh2015single}
M.A. Versteegh, M.E. Reimer, A.A. Berg, G. Juska, V. Dimastrodonato, A.
  Gocalinska, E. Pelucchi and V. Zwiller, Single pairs of time-bin-entangled
  photons. \textit{Physical Review A} \textbf{92}, 033802 (2015).

\bibitem{olislager2010frequency}
L. Olislager, J. Cussey, A.T. Nguyen, P. Emplit, S. Massar, J.M. Merolla and
  K.P. Huy, Frequency-bin entangled photons. \textit{Physical Review A}
  \textbf{82}, 013804 (2010).

\bibitem{zhong201812}
H.S. Zhong, Y. Li, W. Li, L.C. Peng, Z.E. Su, Y. Hu, Y.M. He, X. Ding, W.
  Zhang, H. Li and  others, 12-photon entanglement and scalable scattershot
  boson sampling with optimal entangled-photon pairs from parametric
  down-conversion. \textit{Physical Review Letters} \textbf{121}, 250505
  (2018).

\bibitem{wang201818}
X.L. Wang, Y.H. Luo, H.L. Huang, M.C. Chen, Z.E. Su, C. Liu, C. Chen, W. Li,
  Y.Q. Fang, X. Jiang, J. Zhang, L.N.L. Li, C.Y. Lu and J.W. Pan, 18-qubit
  entanglement with six photons' three degrees of freedom. \textit{Physical
  Review Letters} \textbf{120}, 260502 (2018).

\bibitem{bogdanov267013}
 I.Bogdanov, Graphs with only disjoint perfect matchings.
  \textit{https://mathoverflow.net/q/267013} (2017).

\bibitem{zeilinger1992higher}
A. Zeilinger, M.A. Horne and D.M. Greenberger, Higher Order Quantum
  Entanglement, in Proceedings "Squeezed States and Quantum Uncertainty".
  \textit{National Aeronautics and Space Administration, College Park}
  \textbf{3135}, 73--81 (1992).

\bibitem{bourennane2004experimental}
M. Bourennane, M. Eibl, C. Kurtsiefer, S. Gaertner, H. Weinfurter, O.
  G{\"u}hne, P. Hyllus, D. Bru{\ss}, M. Lewenstein and A. Sanpera, Experimental
  detection of multipartite entanglement using witness operators.
  \textit{Physical Review Letters} \textbf{92}, 087902 (2004).

\bibitem{dur2000three}
W. D{\"u}r, G. Vidal and J.I. Cirac, Three qubits can be entangled in two
  inequivalent ways. \textit{Physical Review A} \textbf{62}, 062314 (2000).

\bibitem{sabidussi1959graph}
G. Sabidussi, Graph multiplication. \textit{Mathematische Zeitschrift}
  \textbf{72}, 446--457 (1959).

\bibitem{weissteingraphproduct}
B. Nicolas and E.W. Weisstein, Graph Product.
  \textit{http://mathworld.wolfram.com/GraphProduct.html} .

\bibitem{weissteinbookgraph}
E.W. Weisstein, Book Graph.
  \textit{http://mathworld.wolfram.com/BookGraph.html} .

\bibitem{drorgraphname294174}
D. Ror., Name and information about this graph.
  \textit{https://mathoverflow.net/questions/294174} (2018).

\bibitem{kiesel2007experimental}
N. Kiesel, C. Schmid, G. T{\'o}th, E. Solano and H. Weinfurter, Experimental
  observation of four-photon entangled Dicke state with high fidelity.
  \textit{Physical Review Letters} \textbf{98}, 063604 (2007).

\bibitem{prevedel2009experimental}
R. Prevedel, G. Cronenberg, M.S. Tame, M. Paternostro, P. Walther, M.S. Kim and
  A. Zeilinger, Experimental realization of Dicke states of up to six qubits
  for multiparty quantum networking. \textit{Physical Review Letters}
  \textbf{103}, 020503 (2009).

\bibitem{wieczorek2009experimental}
W. Wieczorek, R. Krischek, N. Kiesel, P. Michelberger, G. T{\'o}th and H.
  Weinfurter, Experimental entanglement of a six-photon symmetric Dicke state.
  \textit{Physical Review Letters} \textbf{103}, 020504 (2009).

\bibitem{hiesmayr2016observation}
B. Hiesmayr, M. De~Dood and W. L{\"o}ffler, Observation of four-photon orbital
  angular momentum entanglement. \textit{Physical Review Letters} \textbf{116},
  073601 (2016).

\bibitem{gu2019quantum}
X. Gu, M. Erhard, A. Zeilinger and M. Krenn, Quantum experiments and graphs II:
  Quantum interference, computation, and state generation. \textit{Proceedings
  of the National Academy of Sciences} \textbf{116}, 4147--4155 (2019).

\bibitem{huber2013entropy}
M. Huber, M. Perarnau-Llobet and J.I. Vicente, Entropy vector formalism and the
  structure of multidimensional entanglement in multipartite systems.
  \textit{Physical Review A} \textbf{88}, 042328 (2013).

\bibitem{cadney2014inequalities}
J. Cadney, M. Huber, N. Linden and A. Winter, Inequalities for the ranks of
  multipartite quantum states. \textit{Linear Algebra and its Applications}
  \textbf{452}, 153--171 (2014).

\bibitem{scott2004multipartite}
A. Scott, Multipartite entanglement, quantum-error-correcting codes, and
  entangling power of quantum evolutions. \textit{Physical Review A}
  \textbf{69}, 052330 (2004).

\bibitem{goyeneche2014genuinely}
D. Goyeneche and K. {\.Z}yczkowski, Genuinely multipartite entangled states and
  orthogonal arrays. \textit{Physical Review A} \textbf{90}, 022316 (2014).

\bibitem{goyeneche2015absolutely}
D. Goyeneche, D. Alsina, J.I. Latorre, A. Riera and K. {\.Z}yczkowski,
  Absolutely maximally entangled states, combinatorial designs, and
  multiunitary matrices. \textit{Physical Review A} \textbf{92}, 032316 (2015).

\bibitem{fhuberAME35}
F. Huber, IQOQI the Problem35: Existence of absolutely maximally entangled pure
  states.
  \textit{https://oqp.iqoqi.univie.ac.at/existence-of-absolutely-maximally-entangled-pure-states/}
  (2017).

\bibitem{krenn2019questions}
M. Krenn, X. Gu and D. Solt{\'e}sz, Questions on the Structure of Perfect
  Matchings inspired by Quantum Physics. \textit{arXiv preprint
  arXiv:1902.06023} (2019).

\bibitem{raussendorf2001one}
R. Raussendorf and H.J. Briegel, A one-way quantum computer. \textit{Physical
  Review Letters} \textbf{86}, 5188 (2001).

\bibitem{raussendorf2003measurement}
R. Raussendorf, D.E. Browne and H.J. Briegel, Measurement-based quantum
  computation on cluster states. \textit{Physical review A} \textbf{68}, 022312
  (2003).

\bibitem{wootters1998entanglement}
W.K. Wootters, Entanglement of formation of an arbitrary state of two qubits.
  \textit{Physical Review Letters} \textbf{80}, 2245 (1998).

\bibitem{goyeneche2018entanglement}
D. Goyeneche, Z. Raissi, S. Di~Martino and K. {\.Z}yczkowski, Entanglement and
  quantum combinatorial designs. \textit{Physical Review A} \textbf{97}, 062326
  (2018).

\end{thebibliography}

\end{document}